\documentclass[aps,prl,reprint,superscriptaddress,showpacs,showkeys,preprintnumbers,nofootinbib]{revtex4-2}

\usepackage{graphicx}
\usepackage{dcolumn}
\usepackage{bm}
\usepackage{xcolor}
\usepackage{orcidlink}
\usepackage{bbold}
\usepackage{amsmath}
\usepackage{amssymb}

\newcommand{\beq}{\begin{equation}}
\newcommand{\eeq}{\end{equation}}

\newcommand{\beqs}{\begin{eqnarray}}
\newcommand{\eeqs}{\end{eqnarray}}

\newcommand{\dd}{\mbox{d}}
\newcommand{\gs}{g_s}
\newcommand{\ls}{\ell_s}
\newcommand{\Mb}{\overline{M}}

\newcommand{\orcidauthorFAEDO}{0000-0002-3887-2088} 
\newcommand{\orcidauthorPIAI}{0000-0002-2251-0111} 
\newcommand{\orcidauthorRODGERS}{0000-0002-4826-6540} 
\newcommand{\orcidauthorSUBILS}{0000-0003-0104-9722} 
\newcommand{\orcidauthorELANDER}{0000-0001-6348-8021} 

\newcommand{\ccompact}{S$^1_\ell$}

\newcommand{\btriple}{b_0^{\text{\tiny triple}}}
\newcommand{\bCP}{b_0^{\text{\tiny CP}}}
\newcommand{\Lfunction}{{\overline\Lambda}}

\newcommand{\newsec}[1]{\textbf{#1}.}

\usepackage[normalem]{ulem}
\usepackage{soul}
\usepackage{cancel}

\definecolor{mycolor}{RGB}{255,238,140}

\makeatletter 
\renewcommand\onecolumngrid{
\do@columngrid{one}{\@ne}%
\def\set@footnotewidth{\onecolumngrid}
\def\footnoterule{\kern-6pt\hrule width 1.5in\kern6pt}%
}
\renewcommand\twocolumngrid{
        \def\footnoterule{
        \dimen@\skip\footins\divide\dimen@\thr@@
        \kern-\dimen@\hrule width.5in\kern\dimen@}
        \do@columngrid{mlt}{\tw@}
}%
\makeatother    

\begin{document}

\preprint{NORDITA 2025-009}

\title{Light dilaton near critical points in top-down holography}

\author{Daniel Elander\,\orcidlink{\orcidauthorELANDER}}
\email{daniel.elander@gmail.com}
\affiliation{Porto, Portugal}

\author{Ant\'{o}n F. Faedo\,\orcidlink{\orcidauthorFAEDO}}
\email{anton.faedo@uniovi.es}
\affiliation{Departamento de F\'{i}sica, Universidad de Oviedo,  c/ Leopoldo Calvo Sotelo 18, ES-33007, Oviedo, Spain.}
\affiliation{Instituto Universitario de Ciencias y Tecnolog\'{\i}as Espaciales de Asturias (ICTEA), Calle de la Independencia 13, ES-33004, Oviedo, Spain.}

\author{Maurizio Piai\,\orcidlink{\orcidauthorPIAI}}
\email{m.piai@swansea.ac.uk}
\affiliation{Department of Physics, Faculty  of Science and Engineering, Swansea University, Singleton Park, SA2 8PP, Swansea, Wales, UK}

\author{Ronnie Rodgers\,\orcidlink{\orcidauthorRODGERS}}
\email{ronnie.rodgers@su.se}
\affiliation{Nordita, Stockholm University and KTH Royal Institute of Technology, Hannes Alfvéns väg 12, SE-106 91 Stockholm, Sweden.}

\author{Javier~G.~Subils\,\orcidlink{\orcidauthorSUBILS}}
\email{j.gomezsubils@uu.nl}
\affiliation {Institute for Theoretical Physics, Utrecht University, 3584 CC Utrecht, The Netherlands}

\date{\today}

\begin{abstract}
We study a class of UV-complete, strongly coupled, confining three-dimensional field theories, that exhibit a novel stabilisation mechanism for the mass of the lightest scalar composite state, relying on the existence of a critical point. The theories admit a holographic dual description in terms of regular backgrounds in eleven-dimensional supergravity, which retains its rigorous microscopic interpretation in field theory. The phase diagram includes a line of first-order phase transitions ending at the critical point, where the transition becomes of second order. We calculate the mass spectrum of bound states of the field theory, by considering fluctuations around the background solutions, and find that, near the critical point, a hierarchy of scales develops, such that one state becomes parametrically light. We identify this state as the dilaton, the pseudo-Nambu-Goldstone boson associated with the spontaneous breaking of approximate scale invariance, demonstrating the emergence of this composite state in an ab initio calculation that has a  field theory origin. A stabilisation mechanism of this type might be exploited to address hierarchy problems in particle and astroparticle physics.

\end{abstract}

\maketitle

\newsec{Introduction}
\label{Sec:intro}
One promising approach to the electroweak hierarchy problem of the Standard Model is the idea that the Higgs boson~\cite{ATLAS:2012yve,CMS:2012qbp} might be a composite particle---the dilaton---emerging at low energies, in a new, strongly-coupled, confining sector of a more complete theory. In this scenario, the small mass of the Higgs boson, as well as the hierarchy between electroweak and new physics scales, originate from the spontaneous breaking of approximate scale invariance~\cite{Coleman:1985rnk}. The striking, experimentally testable implications of this framework~\cite{Goldberger:2007zk}
motivate phenomenological and effective field theory studies~\cite{Hong:2004td,Dietrich:2005jn,Vecchi:2010gj,Hashimoto:2010nw,DelDebbio:2021xwu,Zwicky:2023fay,Zwicky:2023krx,Eichten:2012qb,Elander:2012fk,Chacko:2012sy,Bellazzini:2012vz,Abe:2012eu,Bellazzini:2013fga,Hernandez-Leon:2017kea}.
A stabilisation mechanism would ensure that explicit symmetry breaking terms are parametrically small,
providing a natural suppression of the dilaton mass, in respect to the strong-coupling scale. 

At a microscopic level, the features of such a scenario are less understood~\cite{Holdom:1986ub,Holdom:1987yu,Appelquist:2010gy,Grinstein:2011dq}. Partly, this is due to the challenge of investigating strongly coupled confining physics. A dilaton is expected to appear in the spectrum of bound states if the confining dynamics is influenced by the proximity, in parameter space, to a weak (second-order) phase transition. Yet, a complete, calculable implementation of these ideas is still missing in four-dimensional gauge theories. In this paper, we take a critical step in this direction, by demonstrating these phenomena in a class of confining theories in lower dimensions.

To do so, we rely on
gauge-gravity dualities~\cite{Maldacena:1997re,Gubser:1998bc,Witten:1998qj,Aharony:1999ti}, which offer an unprecedented opportunity to perform definite computations, linking strongly coupled field theory to weakly coupled gravity in higher dimensions. The physics of confinement in the dual gravity theory is captured by background geometries in which a portion of space shrinks smoothly~\cite{Witten:1998zw,Klebanov:2000hb, Maldacena:2000yy,Butti:2004pk,Nunez:2023xgl,Chatzis:2024kdu,Chatzis:2024top}.
The free energy of different  field-theory configurations is calculable holographically, providing a means to analyse the phase structure. Similarly, the spectrum of bound states is extracted from the fluctuations of the gravity background, treated with the gauge-invariant formalism developed in Refs.~\cite{Bianchi:2003ug,Berg:2005pd,Berg:2006xy,Elander:2009bm,Elander:2010wd,Elander:2014ola,Elander:2018aub,Elander:2020csd}. Explorations of possible backgrounds, looking for evidence of a light dilaton, are found in Refs.~\cite{Goldberger:1999uk,DeWolfe:1999cp,
Goldberger:1999un,Csaki:2000zn,Arkani-Hamed:2000ijo,Rattazzi:2000hs,Kofman:2004tk,
Elander:2009pk,
Elander:2011aa,
Elander:2012yh,Kutasov:2012uq,Evans:2013vca,
Elander:2013jqa,Hoyos:2013gma,Megias:2014iwa,Elander:2015asa,Megias:2015qqh,Athenodorou:2016ndx,
Elander:2017cle,Elander:2017hyr,Elander:2018gte,Pomarol:2019aae,CruzRojas:2023jhw,Pomarol:2023xcc}.

The relation between the nature of phase transitions and the mass of the lightest state in holography has been studied in models chosen on the basis of simplicity arguments (bottom-up holography)~\cite{Elander:2022ebt,Fatemiabhari:2024lct} as well as consistent truncations of fundamental quantum gravity theories (top-down)~\cite{Elander:2020ial,Elander:2020fmv,Elander:2021wkc}. It has been observed that, in proximity to first-order transitions, the mass of the lightest  scalar state in the spectrum is  numerically suppressed, although this typically happens in a branch of metastable states. Yet, in the confining bottom-up model of Ref.~\cite{Faedo:2024zib} (Model B), a line of first-order phase transitions terminates at a critical point. In its close proximity, the mass of the lightest scalar particle can be dialled to be parametrically small in stable states of the theory.

Inspired by this result, but within the context of top-down holography, we present the first example of confining field theories  in which a light dilaton emerges from the dynamics, in the proximity of a critical point at the end of a line of first-order phase transitions. At variance with the model in Ref.~\cite{Faedo:2024zib}, the backgrounds of interest admit a rigorous field-theory dual interpretation, hence filling an important gap in the literature and solving a long standing problem in field theory.
The solutions, constructed within eleven-dimensional supergravity, are regular, based on the circle compactification of those reported in Ref.~\cite{Faedo:2017fbv}, and can be thought of as the double-Wick rotated versions of the solutions constructed in Ref.~\cite{Elander:2020rgv}.
We refer to these  publications for details and provide only the information necessary to make our presentation self-contained.

\newsec{The holographic model}
The truncation detailed in Refs.~\cite{Faedo:2017fbv,Elander:2018gte,Elander:2020rgv} consists of six scalars, \mbox{$\Phi^i=\{ \Phi, U, V, a_J, b_J, b_X\}$}, coupled to gravity in four dimensions. We further compactify it on a circle,  \ccompact, 
where $\ell$ denotes its circumference (for details, see the Supplemental Material). The resulting three-dimensional, classical theory consists of gravity coupled to a sigma model with seven scalars, 
\mbox{$\Phi^a=\{ \Phi^i, \chi \}$}, and a $U(1)$ gauge field, $A_{M}$. The action is
\begin{equation}
\begin{aligned}
\label{eq:action}
{\cal S}_3  &=
\frac{2}{\kappa_3^2} \int \! \dd^3 x \sqrt{|g|} \bigg[\frac{R}{4}
- \frac{1}{2}G_{ab}(\Phi^c) \partial^M \Phi^a \partial_M \Phi^b \\[-1mm]
& \qquad \qquad\qquad\qquad-  \frac{e^{-4\chi}}{16}F_{MN} F^{MN} - \mathcal{V}(\Phi^a)\bigg]\,.
\end{aligned}
\end{equation}
The metric, $g_{MN}$, with spacetime indices $M=0,\,1,\,2$,  has signature mostly `+', determinant $g$, and  Ricci scalar $R$. The field-strength tensor is 
$F_{MN}\equiv \partial_M A_N - \partial_N A_M$, and
scalar field indices are lowered with the sigma-model metric, $G_{ab}$, defined as
\beqs\label{eq:sigma_model_metric}
&&G_{ab}\dd\Phi^a\dd\Phi^b=
\frac{1}{4} \dd\Phi^2+2\dd U^2 + 6\dd V^2 + 4\dd U\dd V  + \dd\chi^2\nonumber\\
&&+2 e^{-4V-\Phi}\dd b_J^2 + 4 e^{-4U-\Phi}\dd b_X^2 + 16 e^{-2U-4V+\frac{\Phi}{2}} \dd a_J^2.
\eeqs
Finally, the potential is $\mathcal V = e^{2\chi} \mathcal V_4(\Phi^i)$ with\footnote{
Compared to Refs.~\cite{Faedo:2017fbv,Elander:2018gte,Elander:2020rgv}, we set $Q_c=0$, following Ref.~\cite{Faedo:2022lxd}.}
\beqs
\label{eq:scalarpotentialV4}
&&\mathcal{V}_4=  8 \left(b_J + b_X\right)^2 e^{-4 U - 8 V - \Phi} -6 e^{-2 U - 6 V} -  2 e^{-4 U - 4 V}\nonumber\\
&&+ \frac{1}{2} e^{-8 V}+ 16 \left[2 a_J +Q_k\left(b_J-b_X\right) -  q_c\right]^2 e^{-6 U - 8 V + \frac{\Phi}{2}} \nonumber\\
&&+ 8 \left(2 a_J -Q_kb_J + q_c\right)^2 e^{-2 U - 12 V + \frac{\Phi}{2}} +Q_k^2 e^{-2 U - 8 V + \frac{3\Phi}{2} }\nonumber\\ 
&&+2Q_k^2 e^{-6 U - 4 V + \frac{3\Phi}{2}}+32\,e^{- 6 U -12 V-\frac{\Phi}{2}} \big[4 a_J \left(b_J + b_X\right) \nonumber\\[-1.5mm]
&&+Q_kb_J\left(b_J-2b_X\right)+2 q_c \left(b_X - b_J\right)\big]^2\,.
\eeqs
The parameters $Q_k$ and $q_c$ appear in the fluxes of the uplifted solutions,
which are dual to three-dimensional \mbox{$\text{U}(N)_ {k}\times\text{U}(N+M)_{-k}$} quiver gauge theories with Chern-Simons interactions at level $k$. With the shorthand notation $\Mb\equiv M-{k}/{2}$, we have (see Ref.~\cite{Faedo:2017fbv})
\begin{equation}\label{gaugeparam}
q_c=\frac{3\pi\ell_s^3g_s}{4}\Mb\,, \qquad  Q_k=\frac{\ell_sg_s}{2}\,k\,,
\end{equation}
with $\ell_s$ and $g_s$ the string length and coupling, respectively. Combining these expressions with the 't~Hooft coupling of the microscopic theory, $\lambda \equiv \ls^ {-1} \gs N$,
the gauge theory depends on a scale, $\Lambda$, which we use to set the scale of dimensionful quantities, and a parameter, $\alpha$:
\begin{equation}\label{eq:units}
    \Lambda = \frac{k^2 \lambda}{6\pi N \overline M}\,, \qquad \alpha = \frac{9 \overline M^3}{256 |k| \pi}\,.
\end{equation}


The system of non-linear equations derived from this action admit a rich space of (non-supersymmetric) background solutions lifting to regular geometries in eleven dimensions—see Refs.~\cite{Faedo:2017fbv,Elander:2020rgv}.
These solutions encapsulate the renormalisation group 
flow
of a family of dual field theories. 
The action and space of solutions are 
rigidly determined, both
by the string theory construction
and by the properties of the dual field theories. 
The solutions can be labelled by two additional parameters:
the circumference,  $\ell$, of the circle, \ccompact, and $b_0 \in (0,\,1)$, which is related to the asymptotic value of the scalar $b_J$---see Eq.~\eqref{eq:UV.expansions2} in the Supplemental Material---and encodes the relative difference in {the (inverse squared) gauge coupling of the two groups in the quiver. In the following, we describe the different phases in the space of these two 
parameters and highlight the key features of the phase transitions that separate them.

\begin{figure}[t]
\begin{center}
\includegraphics{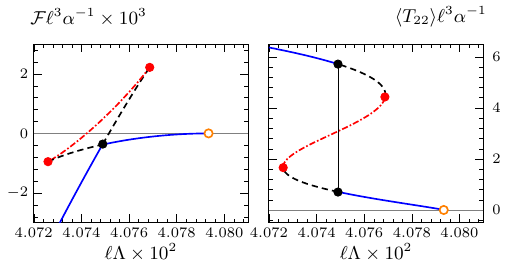}
\caption{%
Free energy density, $\mathcal{F}\ell^3$ (left panel), 
and response function, $\langle T_{22}\rangle \ell^3$ (right panel),
as functions of the circumference, \(\ell\Lambda\), of the compact dimension for confining solutions with 
the representative choice \(b_0 = 0.6836\). The solid black disks indicate the location of a first-order phase transition between two different confining solutions. Beyond the hollow orange disk non-confining solutions are energetically favored.
\label{fig_free_energy_and_T22}
}
\end{center}
\end{figure}

\newsec{Phase diagram and phase transitions}
Each choice of the dimensionless pair $(\ell \Lambda,b_0)$ identifies a solution that is regular when uplifted to eleven-dimensional supergravity and in which the size of the circle \ccompact\, is non-zero everywhere. The dual is a 
three-dimensional \mbox{\textit{non-confining}} state of the field theory, as the quark-antiquark potential is screened~\cite{Faedo:2017fbv}. 
In a region of $(\ell \Lambda,b_0)$ space there are also alternative solutions
for which \ccompact\, shrinks smoothly to zero size, providing the dual 
 to a \textit{confining} state.
With choices of $(\ell \Lambda,b_0)$ for which both solutions exist, the one with the lowest free energy is preferred.

The calculation of the  free energy,  via holographic renormalization~\cite{Bianchi:2001kw, Skenderis:2002wp, Papadimitriou:2004ap} (see the Supplemental Material for the relevant formulae), leads to the identification of  three types of phase transitions. In the range $b_0\in(\bCP,\btriple)\simeq(0.6815,0.6847)$, the free energy is a multi-valued  function of $\ell\Lambda$---see Fig.~\ref{fig_free_energy_and_T22}~(left). A first-order phase transition connects different confining solutions. A discontinuity appears in the response function, $\langle T_{22}\rangle$, the expectation value of the component of the energy-momentum tensor along \ccompact---see the right panel of Fig.~\ref{fig_free_energy_and_T22}. A line of such first-order  transitions exists, represented by the dashed black line in Fig.~\ref{fig_phase_diagram}. As $b_0$ approaches $\bCP$, the transition weakens and eventually disappears into a smooth crossover---see Fig.~\ref{fig_light_dilaton} (left).   

Phase transitions between confining and non-confining states (with zero free energy) also exist, for all values of $b_0\in(0,1)$---see  Figs.~\ref{fig:smallb0phase} and~\ref{fig:largeb0phase}. If $b_0<\btriple$,
the phase transition is such that the free energy of the confining phase touches tangent to the horizontal axes (represented by the hollow orange disk in the top panel of Fig.~\ref{fig_free_energy_and_T22}), as in second order phase transitions. 
However, its true nature is difficult to determine, since close to this particular line, the geometries become increasingly curved.
In contrast, for $b_0>\btriple$ the curve crosses the axes and the phase transition is of first order. All these cases are summarised in the phase diagram in Fig.~\ref{fig_phase_diagram}.

The region of parameter space in proximity to the orange line in Fig. 2 is the only one in which the classical supergravity approximation breaks down. In the rest of the parameter space, quantum gravity and stringy corrections
can be safely neglected, as all curvature invariant are finite, 
and suppressed in the appropriate large-$N$ limit (see Ref.~\cite{Elander:2020rgv} for details).

%
\begin{figure}[t]
\begin{center}
\includegraphics{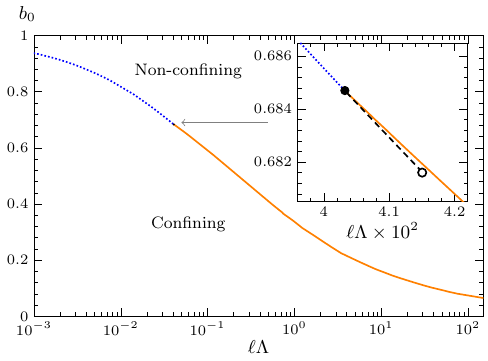}
\caption{%
Phase diagram of the system. Dotted blue and solid orange curves separate confined and non-confined phases, the former identifying first-order phase transitions. Details of the region near the triple point, $\btriple\simeq 0.6847$ (black disk), are shown in the inset panel. A line of first-order phase transitions (dashed black) between different confined states joins $\btriple$  to the critical point at $\bCP\simeq 0.6815$ (white disk).
\label{fig_phase_diagram}
}
\end{center}
\end{figure}

\begin{figure}[!htbp]
\begin{center}
\includegraphics{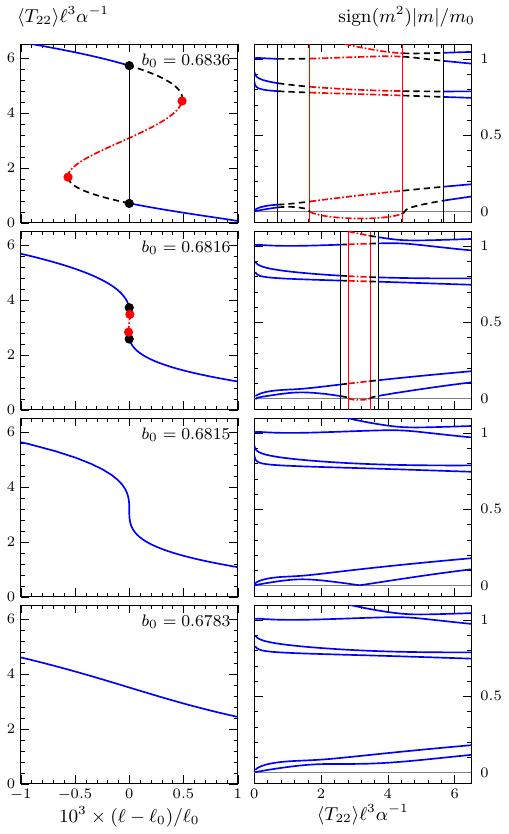}
\caption{%
{
Left panels: component of the stress tensor in the compact direction,  $\langle T_{22}\rangle\ell^3 \alpha^{-1}$,  as a function of its size, \(\ell\),  for representative values of  \(b_0\). For values of \(b_0\) such that there is a first-order phase transition we  define \(\ell_0\) as the value of \(\ell\) at the phase transition. When there is no transition, we instead define \(\ell_0\) as the value of \(\ell\) for which \(\langle T_{22} \rangle \ell^3 \alpha^{-1}\) is the steepest. The range of parameters plotted is chosen to cluster around the line of first-order phase transitions denoted by black dashed lines in Fig.~\ref{fig_free_energy_and_T22}. Right column: mass spectrum of scalar fluctuations, $m$, in units of the lightest vector mass, $m_0$, in the  backgrounds of the left panels.
}
\label{fig_light_dilaton}
}
\end{center}
\end{figure}

\newsec{Mass spectra of fluctuations}
The left panels of Fig.~\ref{fig_light_dilaton} show that the confining solutions are uniquely labelled by
 the pair $(b_0,\,\langle T_{22}\rangle\ell^3 \alpha^{-1})$. We compute the spectrum of small fluctuations of the sigma model coupled to gravity around the confining solutions, by
 exploiting the gauge-invariant approach of Refs.~\cite{Bianchi:2003ug,Berg:2005pd,Berg:2006xy,Elander:2009bm,Elander:2010wd,Elander:2014ola,Elander:2018aub,Elander:2020csd}. The linearised fluctuation equations for the metric, scalar, and vector fields reduce to coupled equations for seven gauge-invariant scalar fluctuations, $\mathfrak{a}^a$, and one vector fluctuation, $\mathfrak{v}$.

The  spectrum of (squared) masses for the fluctuations is obtained by solving the linearised equations,
subject to the requirement that their leading-order modes vanish, both  asymptotically, in the ultraviolet (UV) regime of the field theory, and at the end of space, corresponding to the infrared (IR).

 We numerically determine these masses, $m^2$, using a pseudospectral method~\cite{boyd_book}.
We approximate the solutions as a series of the first \(K\) Chebyshev polynomials of the first kind, and  evaluate the equations and boundary conditions  on the Gauss-Lobatto grid. This approximates the differential equation as a matrix eigenvalue problem.  The eigenvalues, \(m^2\), are extracted using Mathematica's eigenvalue solver. To check convergence of the numerics, we compute the eigenvalues for different values of \(K\),  keeping only those that agree between the computations. More details on our numerical procedure are given in the Supplemental Material.

\begin{figure}
    \includegraphics{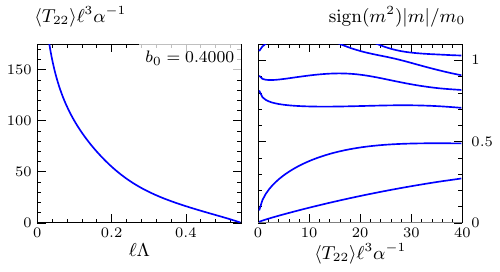}
    \caption{Response function, $\langle T_{22}\rangle\ell^3 \alpha^{-1}$ (left), and  scalar spectrum (right) for a value of \(b_0<\bCP\) chosen well below the critical point. Large values of $\ell \Lambda$ approach the boundary between confining and non-confining configurations.     \label{fig:smallb0phase}}
\end{figure}

\begin{figure}
    \includegraphics{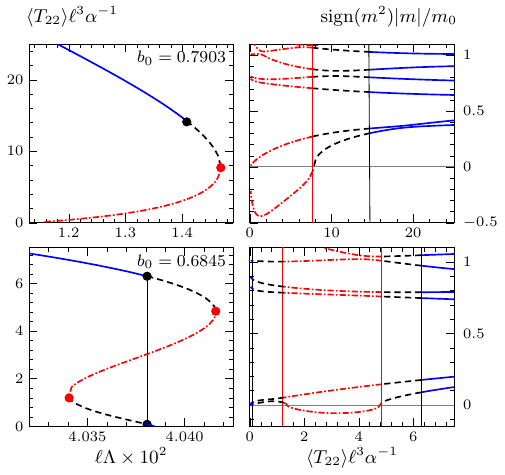}
    \caption{
    Response functions, $\langle T_{22}\rangle\ell^3 \alpha^{-1}$ (left), and scalar spectrum (right) for values of \(b_0>\bCP\), so that a first-order phase transition appears. The top row shows a representative case well above the critical point, where the first-order phase transition is between confining and non-confining solutions. In the bottom row $b_0\simeq\btriple$, slightly below the triple point.
    }
    \label{fig:largeb0phase}
\end{figure}

\newsec{Results}
Our main result is the appearance of a parametrically light scalar, which we identify as a dilaton, in the region of parameter space close to the critical point at the end of a line of first-order phase transitions. This can be seen in Fig.~\ref{fig_light_dilaton}, in which  we show the spectrum obtained for four different values of $b_0\sim \bCP$. 
When a first-order phase transition is present, the spectrum contains a tachyon along the (locally) unstable branch, as seen in the first and second rows of panels. In contrast, when the critical point  is reached (third row), the tachyon disappears, and an exactly massless state is realised. The mass of this state becomes positive and rises as $b_0$ decreases below $\bCP$, when the transition becomes a smooth crossover, as in the last row. Our results thus confirm the statement made in Ref.~\cite{Faedo:2024zib} that a light dilaton is expected near a second-order phase transition, and we explicitly find it in a top-down holographic setup with regular geometry.

\begin{figure}[t]
\begin{center}
\includegraphics[width=.45\textwidth]{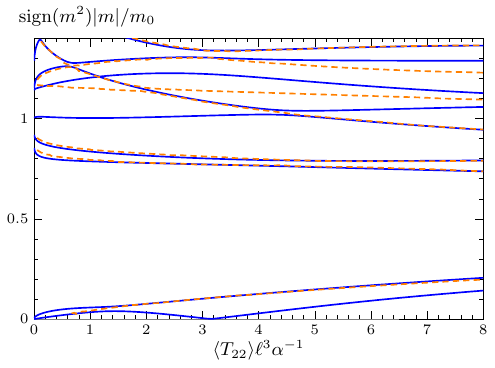}
\caption{%
Mass spectrum of scalar  fluctuations computed for $b_0 = 0.6815$, as in the third row of Fig.~\ref{fig_light_dilaton} (solid blue), and
 in the probe approximation 
(dashed orange), which neglects the effect of the trace of the three-dimensional metric. The probe approximation fails to capture the light mode that appears close to criticality, the dilaton.
\label{fig_probe_approx}
}
\end{center}
\end{figure}

Having demonstrated that 
the presence of a critical point suppresses the mass of the lightest scalar state,
compared to the other scales and masses, we turn attention to its nature.
To this purpose, we repeat the calculation of the spectrum of fluctuations of the confining backgrounds, by introducing a drastic approximation: we ignore the mixing of fluctuations of the sigma-model fields with the trace of the metric, adopting the probe approximation as defined in Ref.~\cite{Elander:2020csd}. As the trace of the metric couples to the trace of the stress-energy tensor of the dual field theory (the dilatation operator), when the probe approximation fails to capture the lightest scalar state, this signals its (approximate) dilaton nature.
We report the results in Fig.~\ref{fig_probe_approx}, for $b_0=\bCP$, as a function of $\langle T_{22}\rangle\ell^3 \alpha^{-1}$.

The probe approximation works well overall, but  fails qualitatively in two important respects.
First, it misses one tower of bound states,  replacing it by a continuum, which we removed 
from the numerical results with appropriate choices of boundary conditions.
Second, it fails to capture  the lightest scalar state, 
in the region of $(b_0,\,\langle T_{22}\rangle\ell^3 \alpha^{-1})$ 
close to criticality, thus demonstrating that this state is an approximate dilaton.

The richness provided by the top-down holographic framework adds additional structure to the spectrum of the theory, compared to the bottom-up model from Ref.~\cite{Faedo:2024zib}. Besides the appearance of a light dilaton near the critical point, the most striking feature is 
the appearance of a second, additional light state, for $b_0\in(0,\btriple)$, when the order parameter, $\left\langle T_{22}\right\rangle$, is small---see Figs.~\ref{fig_light_dilaton} and~\ref{fig:smallb0phase}. As is apparent from Fig.~\ref{fig_probe_approx}, the probe approximation describes well one of these light states, which is hence not related to the dilaton. This region of parameter space corresponds to approaching the solid orange line separating confining and non-confining backgrounds   in Fig.~\ref{fig_phase_diagram}. Unfortunately, when $\left\langle T_{22}\right\rangle \rightarrow  0$, the geometry becomes strongly curved, and eventually singular, preventing us from drawing firm conclusions about this separated, interesting region.

In contrast, for $b_0>\btriple$, the masses of the states  are never suppressed along the stable branch---see Fig.~\ref{fig:largeb0phase}~(top). For completeness, we show in Fig.~\ref{fig:largeb0phase}~(bottom) the spectrum near the special case of the triple point, $b_0\simeq \btriple$, at which the end of the metastable branch touches the first-order phase transition corresponding to the dashed, black line in Fig.~\ref{fig_phase_diagram}.

\newsec{Outlook}
We exhibited a  top-down holographic description of three-dimensional strongly coupled theories in which the mass of a light dilaton state is naturally suppressed by the existence of a nearby critical point in parameter space, putting the results of Ref.~\cite{Faedo:2024zib}  on firm field-theory footing (see also the recent Ref.~\cite{Cresswell-Hogg:2025kvr}, and the lattice study in Ref.~\cite{Lucini:2013wsa}). The emergence of a parametrically light dilaton in a confining theory,  accompanied by a hierarchy of dynamically generated scales,  has been advocated as a way to address hierarchy problems, in the electroweak theory as well as in extensions of the standard model, hence our results  have a range of potential applications.

It would be  interesting to implement this mechanism in four or higher number of dimensions.
If examples exists with the additional feature that other approximate symmetries are also spontaneously broken,
one could further link our findings  with  dilaton effective field theory (dEFT),
in which the dilaton couples to light composite pseudo-Nambu-Goldstone bosons (PNGBs)~\cite{
Matsuzaki:2013eva,Golterman:2016lsd,Kasai:2016ifi,Hansen:2016fri,Golterman:2016cdd,Appelquist:2017wcg,
Appelquist:2017vyy,Cata:2018wzl,Golterman:2018mfm,Cata:2019edh,Appelquist:2019lgk,Golterman:2020tdq,Golterman:2020utm, Appelquist:2022mjb}.
It would also be interesting to investigate possible applications in lower-dimensional field theory, as 
suggested in Ref.~\cite{Cresswell-Hogg:2025kvr}, particularly in the context of conformal perturbation theory---see, e.~g.,  Refs.~\cite{Karananas:2017zrg,Cuomo:2024vfk, Gaberdiel:2008fn,Komargodski:2016auf}.
Applications of dEFT  range from the analysis of special lattice theories~\cite{LSD:2023uzj}, to new dark matter proposals~\cite{Appelquist:2024koa} to
 composite Higgs models in which also the Higgs boson is a PNGB~\cite{Kaplan:1983fs,
Georgi:1984af,
Dugan:1984hq}, as suggested in Refs.~\cite{Appelquist:2020bqj,Appelquist:2022qgl},
by elaborating on Refs.~\cite{Vecchi:2015fma,Ma:2015gra,BuarqueFranzosi:2018eaj}---see the reviews~\cite{Contino:2010rs,
Panico:2015jxa,Witzel:2019jbe, Cacciapaglia:2020kgq,Bennett:2023wjw}
and the tables in Refs.~\cite{Ferretti:2013kya,Ferretti:2016upr,Cacciapaglia:2019bqz}.

The essential finding of the paper is related to the existence of lines of first-order transitions, with end points, and hence a possible strategy for this search would involve applying existing solution-generating techniques, as those discussed in Ref.~\cite{Anabalon:2021tua}, to produce new families of supergravity solutions labelled by more than one parameter.

\newsec{Acknowledgments}
We thank Carlos Hoyos for discussions and collaboration in  early stages of this project. We thank David Mateos for discussions.

A.F. is partially supported by the AEI and the MCIU through the Spanish grant PID2021-123021NB-I00.

The work  MP has been supported by the STFC Consolidated Grant No. ST/T000813/1 and ST/X000648/1.
 MP received funding from the European Research Council (ERC) under the European Union's Horizon 2020 research and innovation program under Grant Agreement No.~813942. 

The work of R.R. was supported by the European Union’s Horizon Europe research and innovation program under Marie Sklodowska-Curie Grant Agreement No.~101104286. Nordita is supported in part by Nordforsk.

J.S. thanks Nordita and Lärkstadens Studiecentrum for their hospitality during his stay in Stockholm from October 9 to 16, 2024, when the main part of this project was undertaken.

{\bf Research Data Access Statement}---The data generated for this manuscript can be downloaded from  
Ref.~\cite{data_release}. 

{\bf Open Access Statement}---For the purpose of open access, the authors have applied a Creative Commons  Attribution (CC BY) licence  to any Author Accepted Manuscript version arising.

\bibliographystyle{apsrev4-1} 
\bibliography{references}

\onecolumngrid
\appendix

\section{SUPPLEMENTAL MATERIAL}

\subsection{Details of the model}

The double-Wick rotated versions of the solutions we study have been analysed extensively in Refs.~\cite{Faedo:2017fbv,Elander:2020rgv,Elander:2018gte}. We include as Supplemental Material details of the model and the relevant background solutions. We also present the relations entering the gauge invariant formalism developed in Ref.~\cite{Bianchi:2003ug,Berg:2005pd,Berg:2006xy,Elander:2009bm,Elander:2010wd,Elander:2014ola}, and discuss the asymptotic behavior of the fluctuations, useful for setting up their boundary conditions.

\subsubsection{Dimensional reduction and background ansatz}
\label{sec:action}

Our starting point is the consistent truncation of type-IIA supergravity (itself related to maximal supergravity in eleven dimensions), obtained after reduction on $\mathbb{CP}^3$, to six scalars and gravity in $D=4$ dimensions, with action given by
\beq
\label{eq:4daction}
{\mathcal S}_4 = \frac{2}{\kappa_4^2}
\int  \dd^4 x  \sqrt{|g_4|} \left[\frac{\mathcal R_4}{4}
-\frac{1}{2} G_{ij}(\Phi^k) \partial^{\hat M}\Phi^i\partial_{\hat M}\Phi^j -\mathcal V_4(\Phi^i)
\right]\,.
\eeq
Here, $g_4$ is the determinant of the four-dimensional metric and $\mathcal R_4$ is the four-dimensional Ricci scalar. The spacetime indices are denoted by $\hat M = 0, \cdots, 3$. The sigma-model metric, $G_{ij}$, is defined in Eq.~\eqref{eq:sigma_model_metric}, 
restricted to the six scalars $\Phi^i = \{ \Phi, U, V, a_J, b_J, b_X\}$, with $i = 1, \cdots, 6$, while the scalar potential, $\mathcal V_4(\Phi^i)$, is given in Eq.~\eqref{eq:scalarpotentialV4}.

We reduce the theory on the circle, \ccompact, parametrised by the coordinate $0 \leq \eta < \ell$. The metric ansatz is
\beq
    \dd s_4^2 = e^{2\chi} \dd s_3^2 + e^{-2\chi} \left( \dd \eta + A_M \dd x^M \right)^2 \,.
\eeq
The resulting reduced action in $D=3$ dimensions is given in Eq.~\eqref{eq:action},  with the identification $\kappa_4^2 = \ell \kappa_3^2$, and consists of gravity, the original six scalars, $\Phi^i$, together with the additional scalar, $\chi$, and gauge field, $A_M$.

We adopt the domain-wall ansatz for the background metric:
\beq
	\dd s^2_3 = \dd r^2 +e^{2A(r)} \eta_{\mu\nu} \dd x^{\mu}\dd x^{\nu} \,,
\eeq
where $r$ is the radial coordinate that we use when writing the equations of motion for the fluctuations.\footnote{The radial coordinate $r$ differs from the one used in Refs.~\cite{Faedo:2017fbv,Elander:2020rgv}.} In the background solutions of interest, the scalars, $\Phi^a(r)$, depend only on $r$, while the vector is set to zero, $A_M=0$.

Solutions to the background equations of motion, descending from the action of Eq.~\eqref{eq:action}, can be obtained by double-Wick rotating the finite-temperature backgrounds constructed in Ref.~\cite{Elander:2020rgv}.\footnote{Compared to Ref.~\cite{Elander:2020rgv}, we make the replacements $(\mathcal{B}_X,\mathcal{B}_J,\mathcal{A}_J) \to (-\mathcal{B}_X,-\mathcal{B}_J,-\mathcal{A}_J)$, needed for regularity, following Ref.~\cite{Faedo:2022lxd}. The equations of motion are invariant under these substitutions.} Such solutions are regular at the end of space in the IR, due to the smooth shrinking of the circle, \ccompact, and implement confinement in the dual field theory, following the same mechanism as in Ref.~\cite{Witten:1998qj}.

In our numerical study, we find it useful to work with a different radial coordinate, $z$, defined by
\begin{equation} \label{eq:r_to_z}
    \dd r = - e^{2 U + 4 V - \frac{\Phi}{2}} \frac{|Q_k|}{u_h z^2} \dd z\,,
\end{equation}
where the parameter $u_h$ is a constant introduced so that $z\in(0,1)$.

\subsubsection{Free energy and response function}

We find it useful to give the expressions for the quantities we plotted in terms of the variables used in Ref.~\cite{Elander:2020rgv}. Consider the functions\footnote{Note that $\Lfunction$ corresponds to $\Lambda$ in previous works. We changed notation to avoid confusion with the energy scale in Eq.~\eqref{eq:units}.
} $f$, $g$, $\Lfunction$, $\mathsf{b}$ and $h$, defined through the following relations:
\begin{equation}
    \begin{aligned}
    e^{2A} &= 256 e^{8 f + 4 g - 4 \Lfunction} \, \mathsf{b} \, h,\quad e^{4U} = 16 e^{4g - \Lfunction} h^{\frac{3}{4}},\quad e^{4V} = 4 e^{4f - \Lfunction} h^{\frac{3}{4}},\quad e^{\Phi} = h^{\frac{1}{4}} e^{\Lfunction},\quad
e^{-2 \chi} = 16 e^{4f + 2g - 2\Lfunction} \mathsf{b}\, h^{\frac{1}{2}}\,,
    \end{aligned}
\end{equation}
and with  a redefinition of the radial coordinate $u \equiv u_h z$. The expansion of these functions near the UV boundary (small $u$) reads
\begin{equation}
\label{eq:UV.expansions1}
    \begin{aligned}
        e^{2f} &= \frac{|Q_k|^2}{2u^2}\left[1-2 u-4 u^2-8 u^3+\left(2 f_4+\frac{77}{4}\right) u^4+\left(-2 f_4+2 f_5+\frac{65}{2}\right)
   u^5\cdots\right]\,,\\
   e^{2g}&= \frac{|Q_k|^2}{4u^2}\left[1-4 u-4 u^2-\left(4 f_4+\frac{109}{2}\right) u^4+\left(14 f_4+2 f_5+\frac{821}{2}\right) u^5+\cdots\right]\,,\\
   e^\Lfunction &=1-4 u^2\cdots\,,\qquad
   \mathsf{b}= 1 + \mathsf{b}_5 u^5 \cdots\,,\qquad h = \frac{4q_c^2}{|Q_k|^6}\frac{16}{15}(1-b_0^2)u^5 +\cdots\,,
    \end{aligned}
\end{equation}
while for the remaining three scalars we have
\begin{equation}
\label{eq:UV.expansions2}
\begin{aligned}
    b_X &= \frac{2q_c}{3|Q_k|}\left(1-b_0-4 b_0 u-12 b_0 u^2\cdots\right)\,,
    \quad
    b_J = \frac{2q_c}{3|Q_k|}\left(
    -1+b_0+4 b_0 u+8 b_0 u^2+\cdots\right)\,,\\
    a_J &= \frac{q_c}{6}\left(-1+2 b_0+16 b_0 u+96 b_0 u^2
    \cdots\right)\,.
\end{aligned}
\end{equation}

In Eqs.~\eqref{eq:UV.expansions1} and \eqref{eq:UV.expansions2}, we explicitly show the free parameters that are fixed by the IR boundary conditions imposed on the background solutions and that appear in the thermodynamic expressions. There are additional free parameters that we are not showing, on which thermodynamic quantities do not depend (see Ref.~\cite{Elander:2020rgv} for complete details). In the IR (small $u - u_h$), all the functions approach a constant value, except $\mathsf{b}$, which vanishes linearly,
\begin{equation}
\begin{array}{rclrclrcl}
 e^{2f} &=& |Q_k|^2 f_h^2  +\cdots  ,
 &e^{2g}&=& |Q_k|^2 g_h^2 + \cdots   ,
 &e^\Lfunction&=& \lambda_h + \cdots ,\\[2mm]
b_J &=& \displaystyle \frac{2q_c}{3|Q_k|} (-1+\xi_h +\cdots),
&\quad b _X&=& \displaystyle \frac{2q_c}{3|Q_k|} (1 - \chi_h + \cdots), 
&\quad a_J&=& \displaystyle\frac{q_c}{6}\left(-1+12 \alpha_h+ \cdots\right), 
 \\[2mm]
h &=& \displaystyle\frac{4q_c^2}{|Q_k|^6} h_h +\cdots & \quad \mathsf{b}&=&  \mathsf{b}_h (u-u_h)+\cdots,&&
\end{array}
\end{equation}
Regularity in the IR fixes the circumference of the circle at the boundary to be
\begin{equation}
    \ell =  - \frac{4 \pi  {h_h^{\frac{1}{2}}}}{\mathsf{b}_h u_h^2} \Lambda^{-1}\,.
\end{equation}
This is interpreted as the inverse temperature of the original solutions, see Ref.~\cite{Elander:2020rgv}. 

With all this information we can write the expressions for the free energy density and the expectation value of the component of the energy-momentum tensor along \ccompact,
\begin{equation}
    \mathcal{F} \ell^3 = - \alpha \frac{32 \pi ^3 h_h^{3/2} }{\mathsf{b}_h^3 u_h^6}\left(3 \mathsf{b}_5-12 f_4-4 f_5-411\right)\, ,\qquad \left\langle T_{22} \right\rangle \ell^3=  \alpha\frac{32 \pi ^3 h_h^{3/2}}{\mathsf{b}_h^3 u_h^6} \left(7 \mathsf{b}_5+12 f_4+4 f_5+411\right)\,.
\end{equation}
The latter corresponds to the energy density of the original finite temperature solutions, before performing the double-Wick rotation. Note that we decided to measure them in units of $\ell$.

\subsubsection{Equations of motion for the fluctuations}
\label{sec:eoms}

In order to compute the spectrum, one fluctuates the metric, the vector, $A_M$, and the scalars, $\Phi^a$, and linearise the equations for small fluctuations around the background. The equation of motion for the transverse, gauge invariant, physical component of the vector, $\mathfrak v$,  is
\beq\label{eq:vectoreom}
	0 = \left[ \partial_r^2 - 4 \partial_r \chi \partial_r - e^{-2A} q^2 \right] \mathfrak v \,,
\eeq
where $q^2 = \eta^{\mu\nu} q_\mu q_\nu= -m^2$, with $q_\mu$ the momentum along the boundary coordinates. The spectrum is given by those values of $m^2$ for which solutions exist that satisfy appropriate boundary conditions in the IR and UV of the geometry. We will discuss the form of these boundary conditions shortly.

Since gravity in $D = 3$ dimensions has no propagating degrees of freedom, there are no spin-2 fluctuations to consider. By contrast, the scalar components of the metric mix non-trivially with the fluctuations of the sigma-model scalars,  $\Phi^a$. By studying gauge-invariant combinations of the fluctuations, the equations can be decoupled (for details regarding the gauge-invariant formalism, see Refs.~\cite{Bianchi:2003ug,Berg:2005pd,Berg:2006xy,Elander:2009bm,Elander:2010wd,Elander:2014ola}). The equations of motion satisfied by the gauge-invariant variable $\mathfrak a^a$ are
\beqs
\label{eq:scalareoms}
	0 &=& \Big[ {\cal D}_r^2 + 2 \partial_{r}{A} {\cal D}_r - e^{-2{A}} q^2 \Big] \mathfrak{a}^a \,\,\nonumber \\
	&& - \Big[  {\mathcal V}^{\,\,a}{}_{\,|c} - \mathcal{R}^a{}_{bcd} \partial_{r}\Phi^b \partial_{r}\Phi^d + 
	\frac{4 (\partial_{r}\Phi^a  {\mathcal V}^{\,b} +  {\mathcal V}^{\,a} 
	\partial_{r}\Phi^b) G_{bc}}{\partial_{r} {A}} + 
	\frac{16  {\mathcal V} \partial_{r}\Phi^a \partial_{r}\Phi^b G_{bc}}{(\partial_{r}{A})^2} \Big] \mathfrak{a}^c\,.
\eeqs
In this expression, sigma-model indices are raised (lowered) with the sigma-model metric $G^{ab}$ ($G_{ab}$), see Eq.~\eqref{eq:sigma_model_metric}.
One then introduces the sigma-model connection, ${\cal G}^a_{\,\,\,\,bc}\equiv\frac{1}{2}G^{ad}\left(\partial_bG_{cd}+\partial_cG_{db}-\partial_dG_{bc}\right)$.
The background covariant derivative is defined as $\mathcal D_r \mathfrak a^a \equiv \partial_r \mathfrak a^a + \mathcal G^a_{\ bc} \partial_r  \Phi^b \mathfrak a^c$. Field-derivatives of the scalar potential are given as $\mathcal V_a \equiv \frac{\partial \mathcal V}{\partial \Phi^a}$ and ${\mathcal V}^a{}_{|b} \equiv \frac{\partial {\mathcal V}^a}{\partial \Phi^b} + \mathcal G^a_{\ bc} {\mathcal V}^c$. The sigma-model Riemann tensor is defined as
$
 {\cal R}^a_{\,\,\,\,bcd}
\equiv \partial_c {\cal G}^a_{\,\,\,\,bd}-\partial_d {\cal G}^a_{\,\,\,\,bc}
+ {\cal G}^e_{\,\,\,\,bd} {\cal G}^a_{\,\,\,\,ce}- {\cal G}^e_{\,\,\,\,bc} {\cal G}^a_{\,\,\,\,de}
$.

\subsubsection{Boundary conditions for the fluctuations}
\label{sec:BCs}

In the process of solving numerically the equations for the fluctuations, we impose boundary conditions in the IR and the UV, such that only subleading modes are kept. This is in line with the accepted prescription in gauge-gravity dualities, that requires that fluctuations be regular and normalisable. For the IR expansions (small $1-z$), all the subleading modes appear at the order of a constant (independent of $z$), which is all we need to know in order to set up the boundary conditions. The UV expansions (small $z$) are more involved. For the vector, $\mathfrak v$, the expansion of the leading fluctuations starts with a constant term, while the subleading mode starts at order $z^5$. For the scalars, $\mathfrak a^a$, the seven leading modes are given by (we only write the very lowest order for each mode):\footnote{In writing these expressions, we have absorbed various factors of the charges into the fluctuations, following Eq.~(3.1) of Ref.~\cite{Elander:2018gte}.}
\beq \label{eq:scalar_leading_modes}
z^{-5}
\left(
\begin{array}{c}
 \frac{12}{5} \\
 1 \\
 1 \\
 0 \\
 0 \\
 0 \\
 0
\end{array}
\right) \,, \quad
z^{-4}
\left(
\begin{array}{c}
 0 \\
 0 \\
 0 \\
 1 \\
 0 \\
 0 \\
 0
\end{array}
\right) \,, \quad
\left(
\begin{array}{c}
 \frac{24b_0}{1-b_0^2} \\
 0 \\
 0 \\
 -\frac{1}{2} \\
 -1 \\
  1 \\
 0
\end{array}
\right) \,, \quad
\left(
\begin{array}{c}
 -20 \\
 1 \\
 1 \\
 0 \\
 0 \\
 0 \\
 0
\end{array}
\right) \,, \quad
\left(
\begin{array}{c}
 0 \\
 0 \\
 0 \\
 0 \\
 0 \\
 0 \\
 1
\end{array}
\right) \,, \quad
z
\left(
\begin{array}{c}
 0 \\
 2 \\
 -1 \\
 0 \\
 0 \\
 0 \\
 0
\end{array}
\right) \,, \quad
z
\left(
\begin{array}{c}
 0 \\
 0 \\
 0 \\
 1 \\
 0 \\
 0 \\
 0
\end{array}
\right) \,.
\eeq
Similarly, the seven subleading modes for $\mathfrak a^a$ are given by:
\beq
z^3
\left(
\begin{array}{c}
 0 \\
 0 \\
 0 \\
 1 \\
 0 \\
 0 \\
 0
\end{array}
\right) \,, \quad
z^4
\left(
\begin{array}{c}
 0 \\
 1 \\
 -\frac{1}{2} \\
 b_0 \\
 0 \\
 0 \\
 0
\end{array}
\right) \,, \quad
z^5
\left(
\begin{array}{c}
 -20 \\
 1 \\
 1 \\
 0 \\
 0 \\
 0 \\
 0
\end{array}
\right) \,, \quad
z^5
\left(
\begin{array}{c}
 0 \\
 0 \\
 0 \\
 0 \\
 0 \\
 0 \\
 1
\end{array}
\right) \,, \quad
z^6
\left(
\begin{array}{c}
 \frac{255}{11} \\
 \frac{249}{44} \\
 -\frac{147}{44} \\
 \frac{8 b_0}{5} \\
 \frac{b_0^2-1}{b_0} \\
 \frac{1-b_0^2}{b_0} \\
 0
\end{array}
\right) \,, \quad
z^8
\left(
\begin{array}{c}
 0 \\
 0 \\
 0 \\
 1 \\
 0 \\
 0 \\
 0
\end{array}
\right) \,, \quad
z^{10}
\left(
\begin{array}{c}
 \frac{12}{5} \\
 1 \\
 1 \\
 0 \\
 0 \\
 0 \\
 0
\end{array}
\right) \,.
\eeq
This information is important to find the spectrum numerically, as  knowing at which power the subleading modes appear for the first time  allows to impose  the appropriate boundary conditions for the pseudospectral problem, as we will see later.

\subsubsection{Probe approximation}
\label{sec:probes}

In the probe approximation, one neglects the mixing between the metric and the scalar fluctuations. This corresponds to removing the two last terms in Eq.~\eqref{eq:scalareoms}, resulting in the equations of motion for the probe scalars,~$\mathfrak p^a$:
\beqs\label{eq:scalareoms_probe}
0&=&\left[\frac{}{}{\cal D}_r^2 + 2 \partial_r A\,{\cal D}_r -e^{-2A} q^2\right]\mathfrak{p}^a-\left[\frac{}{}V^a_{\,\,\,\,|c}-{\cal R}^a_{\,\,\,\,bcd}\partial_r\Phi^b\partial_r \Phi^d
\right]\mathfrak{p}^c \,.
\label{eq:probe}
\eeqs
In the present case, this modifies so drastically the UV behaviour of the fluctuations as to introduce oscillatory modes, signalling a continuum in the spectrum. More precisely, the leading modes for the probe scalars, $\mathfrak p^a$, are given by
\beq
z^{-5}
\left(
\begin{array}{c}
 \frac{12}{5} \\
 1 \\
 1 \\
 0 \\
 0 \\
 0 \\
 0
\end{array}
\right) \,, \quad
z^{-4}
\left(
\begin{array}{c}
 0 \\
 0 \\
 0 \\
 1 \\
 0 \\
 0 \\
 0
\end{array}
\right) \,, \quad
\left(
\begin{array}{c}
 \frac{24b_0}{1-b_0^2} \\
 0 \\
 0 \\
 - \frac{1}{2} \\
 - 1 \\
 1 \\
 0
\end{array}
\right) \,, \quad
\left(
\begin{array}{c}
 -20 \\
 1 \\
 1 \\
 0 \\
 0 \\
 0 \\
 4
\end{array}
\right) \,, \quad
z
\left(
\begin{array}{c}
 0 \\
 2 \\
 -1 \\
 0 \\
 0 \\
 0 \\
 0
\end{array}
\right) \,, \quad
z
\left(
\begin{array}{c}
 0 \\
 0 \\
 0 \\
 1 \\
 0 \\
 0 \\
 0
\end{array}
\right) \,,
\eeq
while the subleading modes are
\beq \label{eq:probe_subleading}
z^3
\left(
\begin{array}{c}
 0 \\
 0 \\
 0 \\
 1 \\
 0 \\
 0 \\
 0
\end{array}
\right) \,, \quad
z^4
\left(
\begin{array}{c}
 0 \\
 1 \\
 -\frac{1}{2} \\
 b_0 \\
 0 \\
 0 \\
 0
\end{array}
\right) \,, \quad
z^5
\left(
\begin{array}{c}
 -20 \\
 1 \\
 1 \\
 0 \\
 0 \\
 0 \\
 4
\end{array}
\right) \,, \quad
z^6
\left(
\begin{array}{c}
 \frac{255}{11} \\
 \frac{249}{44} \\
 -\frac{147}{44} \\
 \frac{8 b_0}{5} \\
 \frac{b_0^2-1}{b_0} \\
 \frac{1-b_0^2}{b_0} \\
 0
\end{array}
\right) \,, \quad
z^8
\left(
\begin{array}{c}
 0 \\
 0 \\
 0 \\
 1 \\
 0 \\
 0 \\
 0
\end{array}
\right) \,, \quad
z^{10}
\left(
\begin{array}{c}
 \frac{12}{5} \\
 1 \\
 1 \\
 0 \\
 0 \\
 0 \\
 0
\end{array}
\right) \,.
\eeq
Additionally, there are two oscillatory modes, $\mathfrak p^a = z^{\frac{1}{2}\left(5 \pm i \sqrt{55}\right)} \left(-20, 1, 1, 0, 0, 0, -28\right)^T$.

We impose boundary conditions chosen to set to zero the leading modes as well as both oscillatory modes, hence reducing the space of solutions to the discrete part of the spectrum, which we hence expect to contain six towers of states. We interpret our results as follows. First, the states corresponding to masses that are well approximated by the probe calculation, do not mix heavily with the metric and hence are non-dilatonic. Second, those states for which the result of the probe calculation deviates significantly from the correct calculation, contain a non-trivial dilatonic component. The very existence of oscillatory modes, and the resulting spurious continuum, are an artefact of having neglected the 
non-trivial effects of mixing with the metric.

\subsection{Numerical procedure for computation of the spectrum}

In this section we provide further details on the pseudospectral method that we use to compute the fluctuation spectra. The equations of motion that we need to solve for the fluctuations, Eqs.~\eqref{eq:vectoreom} and \eqref{eq:scalareoms}, can be rearranged to take the form
\begin{subequations}\label{eq:gauge_invariant_fluctuation_equations} 
\begin{eqnarray} 
     C(z) \partial_z^2 \mathfrak{a}^a{}(z) + D^a{}_b(z) \partial_z \mathfrak{a}^b(z) + E^a{}_b(z) \mathfrak{a}^b(z) &= m^2 \mathfrak{a}^a(z)\,,\quad
     \label{eq:scalar_fluctuation_equations} 
     \\
    \tilde{C}(z) \partial_z^2 \mathfrak{v}(z) +\tilde{D}(z) \partial \mathfrak{v}(z) +\tilde{E}(z)  \mathfrak{v}(z) &= m^2 \mathfrak{v}(z)\,,\quad
\end{eqnarray}
\end{subequations}
where $C(z)$, $\tilde{C}(z)$, $D^a{}_b(z)$, $\tilde{D}(z)$, $E^a{}_b(z)$, and $\tilde{E}(z)$ are known functions that depend on the background fields and their derivatives.  
In the IR, at \(z=1\), the space ends.  This is a regular singular point for Eqs.~\eqref{eq:gauge_invariant_fluctuation_equations}; $C(z)$ and $\tilde{C}(z)$ vanish linearly at \(z=1\), while the remaining coefficient functions remain finite. Imposing as boundary conditions the requirement that the fluctuations are regular at \(z=1\), Eq.~\eqref{eq:gauge_invariant_fluctuation_equations} then provides a set of linear relations between the fluctuations and their first derivatives, evaluated at \(z=1\).

Conversely,  the  UV 
boundary, at \(z=0\), is an irregular singular point. For the scalar fluctuations, the solutions we want to find are such that the coefficients of the leading modes in Eq.~\eqref{eq:scalar_leading_modes} vanish. An efficient way to impose this condition is to define new fluctuations, \(\bar{\mathfrak{a}}^a(z)\), related to the unbarred fluctuations by a suitable power of $z$, such that they vanish linearly at the boundary if the boundary conditions are satisfied, \(\bar{\mathfrak{a}}(0) = 0\). We also impose the boundary condition \(\mathfrak{v}(0) =0\) on the vector fluctuations.

To determine the spectrum numerically, we use a pseudospectral method~\cite{boyd_book}. We describe in some detail here the procedure for the scalar fluctuations, \(\bar{\mathfrak{a}}^a\); the treatment of the vector fluctuation proceeds along the same lines. The first step is to approximate the fluctuations as linear combinations  of the first \(K\) Chebyshev polynomials of the first kind, \(T_k\):
\begin{equation} \label{eq:chebyshev_expansion}
    \bar{\mathfrak{a}}^a(z) = \sum_{k=0}^{K-1} c_k^a T_k(2z-1) \, ,
\end{equation}
for some appropriate chocie of coefficients, \(c_k^a\). We then evaluate the fluctuation equation, Eq.~\eqref{eq:scalar_fluctuation_equations}, on the \(K\)-point Gauss-Lobatto grid linearly mapped to the range \(z \in [0,1]\),
\begin{equation}
    z_k \equiv \sin^2 \left( \frac{\pi k}{2(K-1)}\right)\, ,
    \qquad
    k = 0, \cdots, K-1 \, ,
\end{equation}
where at \(z_0 = 0\) we use the boundary conditions \(\bar{\mathfrak{a}}^a(0)=0\), in place of the equation of motion. When \(\bar{\mathfrak{a}}^a\) takes the form in Eq.~\eqref{eq:chebyshev_expansion}, differentiation identities of the Chebyshev polynomials are used to express \(\bar{\mathfrak{a}}'(z_k)\) and \(\bar{\mathfrak{a}}''(z_k)\) as linear combinations of \(\bar{\mathfrak{a}}^a(z_j)\), evaluated at the other grid points. Thus, this procedure approximates Eq.~\eqref{eq:scalar_fluctuation_equations} as a generalised matrix eigenvalue problem,
\begin{equation} \label{eq:fluctuation_eigenvalue_problem}
    A |\bar{\mathfrak{a}}\rangle = m^2 B|\bar{\mathfrak{a}}\rangle \, ,
\end{equation}
for \(7K \times 7K\)  matrices, \(A\)  and \(B\), where the eigenvectors, \(| \bar{\mathfrak{a}} \rangle\) , contain the fluctuations evaluated on the grid points,
\begin{equation}
    | \bar{\mathfrak{a}} \rangle = \left(\bar{\mathfrak{a}}^1(z_0), 
    \cdots,
    \bar{\mathfrak{a}}^7(z_0),
    \bar{\mathfrak{a}}^1(z_1),
    \cdots,
    \bar{\mathfrak{a}}^7(z_1),
    \cdots,
    \bar{\mathfrak{a}}^1(z_K), 
    \cdots,
    \bar{\mathfrak{a}}^7(z_K)\right)\, .
\end{equation}

The numerical approximation to the spectrum is then obtained as the set of eigenvalues of Eq.~\eqref{eq:fluctuation_eigenvalue_problem}, which we compute using Mathematica's eigenvalue solver. For a grid size \(K\), only around half of the eigenvalues are expected to provide approximations to true modes, the remaining eigenvalues being spurious~\cite{boyd_book}. The values of the spurious modes depend on \(K\), so we eliminate them by computing the sectrum for two different grid sizes, \(K_1\) and \(K_2\), keeping only those eigenvalues that agree between the two computations, using an algorithm that we borrow from Ref.~\cite{boyd_book}. Most of the eigenvalues plotted in Figs.~\ref{fig_light_dilaton},~\ref{fig:smallb0phase}, and~\ref{fig:largeb0phase} have been computed by using grid sizes \((K_1,K_2) = (90,100)\).

For small values of the response function, \(\langle T_{22} \rangle \ell^3 \alpha^{-1} \lesssim 1\), we found that the eigenvalues, particularly the lightest modes, become numerically unstable when using the simple algorithm described so far. The problem appears to be that the eigenfunctions \(\bar{\mathfrak{a}}^a(z)\) corresponding to these eigenvalues develop large gradients near \(z=1\), and is likely connected to the fact that the curvature of the background solution grows large in this regime. To remedy this problem, we found it useful to define a new radial coordinate via the relation
\begin{equation} \label{eq:z_bar}
    \bar{z} = 1 - \sqrt{1-z}\,,
\end{equation}
chosen so that \(1-z = (1- \bar{z})^2\). The first derivatives of the fluctuations with respect to \(\bar{z}\) then vanish near \(\bar{z}=1\). Combining this coordinate transformation with an increase of grid sizes to \((K_1,K_2) = (190,200)\), we have been able to obtain reliable numerical determinations of the spectrum for smaller values of \(\langle T_{22} \rangle \ell^3 \alpha^{-1}\).

The computation of the probe-approximation spectrum from Eq.~\eqref{eq:scalareoms_probe} has been carried out using the same pseudospectral method. The existence of the oscillatory modes mentioned after Eq.~\eqref{eq:probe_subleading} introduces an extra complication. Although we impose boundary conditions such that the oscillatory modes are turned off, the fact that they could in principle be present introduces many additional spurious modes to the spectrum computed with a given grid size. To handle this undesirable feature, we found it useful to work with the radial coordinate \(\bar{z}\) defined in Eq.~\eqref{eq:z_bar} throughout, and compute the spectra for five different grid sizes, keeping only those modes that agree between at least three different grid sizes. The probe spectrum in Fig.~\ref{fig_probe_approx} has been  computed using grid sizes (120,140,150,160,180).

\end{document}